\documentclass[11pt]{article}
\usepackage{etex} %
\usepackage{natbib}

\usepackage{comment}
\usepackage{amsthm,amsmath,amssymb,amsfonts,latexsym,stmaryrd}
\usepackage[all]{xy}
\usepackage{euscript}
\usepackage{graphicx}
\usepackage{epsfig}
\usepackage{tikz,ifdraft}
\usetikzlibrary{arrows,positioning,matrix,backgrounds,calc,fit,decorations.pathreplacing} 
\usepackage{nicefrac}

\usepackage{eurosym}

\theoremstyle{plain}
\newtheorem{theorem}{Theorem}[section]

\newtheorem{proposition}[theorem]{Proposition}

\theoremstyle{definition}  %

\newcommand{\beqa}{\begin{eqnarray*}}
\newcommand{\eeqa}{\end{eqnarray*}\par\noindent}

\newcommand{\vsa}{\vspace{.1in}}

\newcommand{\id}{\mathsf{id}}
\renewcommand{\emph}[1]{\textbf{#1}}

\newcommand{\pow}{\mathcal{P}}

\newcommand{\ket}[1]{{|} #1\rangle}
\newcommand{\ie}{\textit{i.e.}~}

\newcommand{\Set}{\mathbf{Set}}

\newcommand{\vn}{\varnothing}

\newcommand{\lrarr}{\leftrightarrow}

\newcommand{\IMP}{\; \rightarrow \;}

\newcommand{\da}{{\downarrow}}

\newcommand{\EM}{\mathsf{EM}}

\newcommand{\op}{\mathsf{op}}

\newcommand{\DD}{\mathcal{D}}
\newcommand{\Real}{\mathbb{R}}

\newcommand{\Bool}{\mathbb{B}}

\newcommand{\ua}{{\uparrow}}

\newcommand{\prob}{\mathsf{Prob}}
\newcommand{\vphi}{\varphi}

\newcommand{\SQT}{\!\sqrt{2}}

\newcommand{\rsqa}{\, \leadsto \,}

\def\Mdot{\text{ .}}

\def\cat#1{\mathbf{\mathsf{#1}}}

\renewcommand{\Set}      {\cat{Set}        }  %

\def\op{\mathsf{op}}

\usepackage{colonequals}

\newcommand{\adjointfdec}[4]{(#1 \vdash #2)\colon \xymatrix{#3 \ar@<.5ex>[r]^{f_*} & \ar@<.5ex>[l]^{f^*} #4}}

\newcommand{\Cont}{\mathsf{Cont}}
\newcommand{\Feat}{\mathsf{Feat}}
\newdir{ >}{{}*!/-7.5pt/@{>}}

\newcommand{\Rgeq}{\Real_{\geq 0}}
\newcommand{\hb}{\bar{h}}

\newcommand{\rTo}{\to}

\usepackage[margin=1.25in]{geometry}

\begin{document}

\author{Samson Abramsky \\
%\address{Department of Computer Science, University of Oxford}
%\email{\{samson\,|\,ruibar\,|\,kishida\,|\,rayl\,|\,shaman\}@cs.ox.ac.uk}
%\email{\{samson.abramsky\,|\,rui.soares.barbosa\,|\,kohei.kishida\,|\,raymond.lal\,|\,shane.mansfield\}@cs.ox.ac.uk}
%\institute{
Department of Computer Science, University of Oxford \\
\indent
\textup{\texttt{samson.abramsky@cs.ox.ac.uk}}}

\title{Contextuality: At the Borders of Paradox}

\maketitle

%\chapter{Contextuality: At the Borders of Paradox \\ {\normalsize Samson Abramsky}}

\begin{abstract}
Contextuality is a key feature of quantum mechanics.
We present the sheaf-theoretic approach to contextuality introduced by Abramsky and Brandenburger, and show how it covers a range of logical and physical phenomena ``at the borders of paradox''.
\end{abstract}

\section{Introduction}\label{sec:intro}

Logical consistency is usually regarded as a minimal requirement for scientific theories, or indeed for rational thought in general. Paraconsistent logics \citep{priest2002paraconsistent} aim at constraining logical inference, to prevent inconsistency leading to triviality.

However, a richer, more nuanced view of consistency is forced on us if we are to make sense of one of the key features of quantum physics, namely \emph{contextuality}.

The phenomenon of contextuality is manifested in classic No-Go theorems such as the Kochen-Specker paradox \citep{kochen:67}; and in a particular form, also appears in Bell's theorem \citep{bell:64}, and other non-locality results such as the Hardy paradox \citep{hardy:93}. The close relationship of these results to issues of consistency is suggested by the very fact that terms such as ``Kochen-Specker paradox'' and ``Hardy paradox'' are standardly used.

At the same time, these arguments are empirically grounded. In fact, a  number of experiments to test for contextuality have already been performed \citep{bartosik:09,kirchmair:09,zu:12}. Moreover, it has been argued that contextuality can be seen as an essential \emph{resource} for quantum advantage in computation and other information-processing tasks \citep{howard:14,PhysRevA.88.022322}.

What, then, is the essence of contextuality? In broad terms, we propose to describe it as follows:

\begin{quote}
Contextuality arises where we have a family of data which is \emph{locally consistent},\\ 
but \emph{globally inconsistent}.
\end{quote}

In more precise terms, suppose that we have a family of data $\{ D_c \}_{c \in \Cont}$, where each $D_c$ describes the data which is observed in the context $c$. We can think of these contexts as ranging over various experimental or observational scenarios $\Cont$. This data is locally consistent in the sense that, for all contexts $c$ and $d$, $D_c$ and $D_d$ agree on their overlap; that is, they give consistent information on those features which are common to both contexts.

However, this data is globally inconsistent if there is no global description $D_g$ of \emph{all} the features which can be observed in \emph{any} context, which is consistent with all the local data $D_c$, as $c$ ranges over the set of possible contexts $\Cont$.

An immediate impression of how this situation might arise is given by impossible figures such as the Penrose tribar \citep{penrose1992cohomology}, shown in Figure~1.
\begin{figure}
\begin{center}
\begin{tikzpicture}[scale=0.8]
  \path (.5,.866) coordinate (L1);
  \path (1,0) coordinate (L2);
  \path (2.5,.866) coordinate (L3);
  \path (3,1.732) coordinate (L4);
  \path (4,1.732) coordinate (R1);
  \path (5,1.732) coordinate (R2);
  \path (6,0) coordinate (R3);
  \path (6.5,.866) coordinate (R4);
  \path (3,3.464) coordinate (T1);
  \path (3,5.196) coordinate (T2);
  \path (3.5,2.598) coordinate (T3);
  \path (4,5.196) coordinate (T4);
  \begin{scope}[black]
  \filldraw[fill=gray] (L1) -- (T2) -- (R2) -- (R1) -- (T1) -- (L2) -- cycle;
  \filldraw[fill=black] (L2) -- (T1) -- (T3) -- (L3) -- (R4) -- (R3) -- cycle;
  \draw (L3) -- (R4) -- (T4) -- (T2) -- (R2) -- (L4) -- cycle;
  \end{scope}
\end{tikzpicture}
\end{center}
\caption{Penrose tribar}
\label{fig:tribar}
\end{figure}
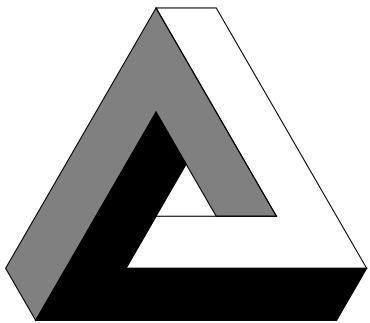%

If we take each leg of the tribar, and the way each pair of adjacent legs are joined to each other, this gives a family of locally consistent data, where consistency here refers to realizability as a solid object in 3-space. However, the figure as a whole is inconsistent in this sense.
We will see more significant examples arising from quantum mechanics shortly.

Why should we regard this phenomenon of contextuality as disturbing?
Consider the following situation. Certain fundamental physical quantities are being measured, e.g.~electron spin or photon polarization. Some of these physical quantities can be measured together.
In a famous example, Alice and Bob are spacelike separated, and a pair of particles are prepared at some source, and one is sent to Alice, and one to Bob. Then they can each measure the spin of their particle in a given direction. Each such measurement has two possible outcomes, spin up or down in the given direction. We shall refer to this choice of direction as a \emph{measurement setting}. Now imagine this procedure being repeated, with Alice and Bob able to make different choices of measurement setting  on different repetitions. This gives rise to a family of local data, the statistics of the outcomes they observe for their measurements. This set of local data is moreover locally consistent, in the sense that Alice's observed statistics for the outcome for various measurement settings is independent of Bob's choice of measurements, and vice versa.
However, let us suppose that this data is globally inconsistent: there is no way of explaining what Alice and Bob see in terms of some joint distribution on the outcomes for all possible measurement settings.

The import of Bell's theorem and related results  is that exactly this situation is predicted by quantum mechanics. Moreover, these predictions have been extensively confirmed by experiment, with several loophole-free Bell tests having been published in 2015 \citep{hensen2015loophole,shalm2015strong,giustina2015significant}.

What this is saying is that the fundamental physical quantities we are measuring, according to our most accurate and well-confirmed physical theory, cannot be taken to have objective values which are independent of the context in which they are being measured.

In the light of this discussion, we can elucidate the connection between contextuality and inconsistency and paradox. There is no inconsistency in what we can actually observe directly --- no conflict between logic and experience. The reason for this is that we cannot observe all the possible features of the system at the same time. In terms of the Alice-Bob scenario, they must each choose one of their measurement settings each time they perform a measurement. Thus we can never observe directly what the values for the other measurement settings are, or ``would have been''. In fact, we must conclude that they have \emph{no well-defined values}.

The key ingedient of quantum mechanics which enables the possibility of contextual phenomena, while still allowing a consistent description of our actual empirical observations, is the presence of \emph{incompatible observables}; variables that cannot be measured jointly. In fact, we can say that the experimental verification of contextual phenomena shows that \emph{any} theory capable of predicting what we actually observe must incorporate this idea of incompatible observables.

The compatible families of observables, those which \emph{can} be measured together, provide the empirically accessible windows onto the behaviour of microphysical systems. These windows yield local, and locally consistent, information. However, in general there may be no way of piecing this local data together into a global, context-independent description. This marks a sharp difference with classical physics, with consequences which are still a matter for foundational debate.

The phenomenon of contextuality thus leaves quantum mechanics, and indeed any empirically adequate ``post-quantum'' theory, on the borders of paradox, without actually crossing those borders. This delicate balance is not only conceptually challenging, but appears to be deeply implicated in the \emph{quantum advantage}; the possibility apparently offered by the use of quantum resources to perform various information-processing tasks better than can be achieved classically. 
%The borders of paradox are a zone where interesting logical and informatic phenomena emerge; an analogy with the edge of chaos suggests itself.

\section{Basic Formalization}

We shall now show how our intuitive account of contextuality can be formalized, using tools from category theory. This will provide the basis for a mathematical theory of contextuality.
It can also, perhaps,  serve as a case study for how conceptual discussions can be turned into precise mathematics using categories.

We recall the basic ingredients of our informal discussion:
\begin{itemize}
\item Notions of \emph{context} of observation or measurement, and of \emph{features} of the system being observed or measured.
Certain features, but in general not all, can be observed in each context. Those features which can be observed in the same context are deemed \emph{compatible}.

\item A collection of observations will give rise to a family of data $\{ D_c \}$ describing the information yielded by these observations, ranging over the contexts.

\item We say that such a family of data is \emph{locally consistent} if for each pair of contexts $c$, $d$, the data $D_c$, $D_d$ yield consistent descriptions of the features which are common to $c$ and $d$.

\item We say that the family of data is \emph{globally consistent} if there is some data description $D_g$ giving information on all the features which can appear in any context, and which is consistent with the local data, in the sense that each local description $D_c$ can be recovered by restricting $D_g$ to the features which can be observed in the context $c$.
If there is no such global description, then we say that the family of data $\{ D_c \}$ is \emph{contextual}.
\end{itemize}
To formalize these ideas, we fix sets $\Cont$, $\Feat$ of contexts and features, together with a function 
\[ \Phi : \Cont \rTo \pow(\Feat) \]
which for each context $c$ gives the set of features which can be observed in $c$.

We shall make the assumption that this function is injective, ruling out the possibility of distinct contexts with exactly the same associated features. This is not essential, but simplifies the notation, and loses little by way of examples.
We can then \emph{identify} contexts with their associated set of features, so we have $\Cont \subseteq \pow(\Feat)$.
We moreover assume that this family is closed under subsets: if $C \subseteq D \in \Cont$, then $C \in \Cont$. This says that if a set of features is compatible, and can be observed jointly, so is any subset.

To model the idea that there is a set of possible data descriptions which can arise from  performing measurements or observations on the features in a context, we assume that there is a map
\[ P : \Cont \rTo \Set \]
where $P(C)$ is the set of possible data for the context $C$.

We now come to a crucial point. In order to define both local and global consistency, we need to make precise the idea that the information from a larger context, involving more features, can be cut down or restricted to information on a smaller one. That is, when $C \subseteq D$, $C, D \in \Cont$, we require  a function
\[ \rho^D_C : P(D) \rTo P(C) . \]
We call such a function a \emph{restriction map}. We require that these functions satisfy the obvious conditions:
\begin{itemize}
\item If $C \subseteq D \subseteq E$, then $\rho^D_C \circ \rho^E_D = \rho^E_C$
\item $\rho^C_C = \id_{P(C)}$.
\end{itemize}
This says that $P$ is a functor, 
\[ P : \Cont^{\op} \rTo \Set \]
where $\Cont$ is the poset \textit{qua} category of subsets under inclusion.
Such a functor is called a \emph{presheaf}.

In fact, it will be convenient to assume that $P$ is defined on arbitary subsets of features, compatible or not, so that $P$ is a presheaf
\[ P : \pow(\Feat)^{\op} \rTo \Set \]
This will allow us to formulate global consistency.

\textbf{Notation} It will be convenient to write $d |_C := \rho^D_C(d)$ when $d \in P(D)$ and $C \subseteq D$.
In this notation, the functoriality conditions become:
\[ (d |_D) |_C = d |_C, \qquad d |_C = d. \]

We now have exactly the tools we need to define local and global consistency.

A family of data is now a family $\{ d_C \}_{C \in \Cont}$, where $d_C \in P(C)$. Such a family is locally consistent if whenever $C \subseteq D$, then $d_C = (d_D) |_C$.
Equivalently, for any $C, D \in \Cont$, $d_C |_{C \cap D} = d_D |_{C \cap D}$. This formalizes the idea that the data from the two contexts yield consistent information with respect to their overlap, \ie their common features.

The family is globally consistent if there is some $d_g \in P(\Feat)$, global data on the whole set of features, such that, for all contexts $C$, $d_g |_C = d_C$; that is, the local data can be recovered from the global description. It is the absence of such a global description which is the signature of contextuality.

These notions of local and global consistency have significance in the setting of sheaf theory \citep{maclane:92,kashiwara2005categories}, which is precisely the mathematics of the passage between local and global descriptions, and the obstructions to such a passage which may arise.
If we think of the family of subsets $\Cont$ as an ``open cover'', in the discrete topology $\pow(\Feat)$, then a locally consistent family $\{ d_C \}_{C \in \Cont}$ is exactly a \emph{compatible family} in the terminology of sheaf theory. The condition for this family to be globally consistent is exactly that is satisfies the \emph{gluing condition}. The condition for the presheaf $P$ to be a sheaf is exactly that for every open cover, every compatible family satisfies the gluing condition in a unique fashion.
Thus contextuality arises from obstructions to instances of the sheaf condition.

We will return to this link to the topological language of sheaf theory shortly.

\section{Example}

We shall now examine an Alice-Bob scenario, of the kind discussed in Section~1, in detail. The scenario is depicted in Figure~2.
Alice can choose measurement settings $a_1$ or $a_2$, while Bob can choose $b_1$ or $b_2$. Alice can measure her part of the system with her chosen setting, and observe the outcome $0$ or $1$. Bob can perform the same operations with respect to his part of the system.
They send the outcomes  to a common target.

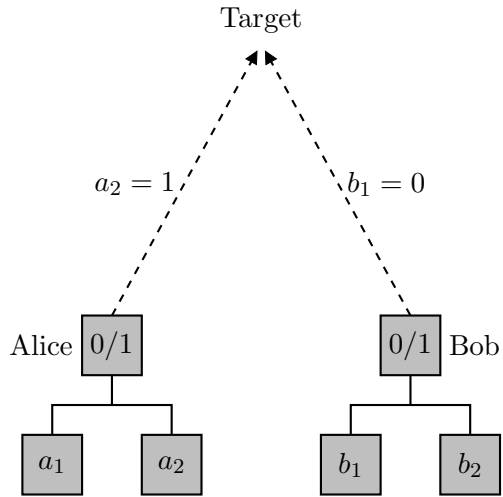
\begin{figure}
\label{bitreg}
\begin{center}
\begin{tikzpicture}[scale=2.54]
% dpic version 2011.03.17 option -g for TikZ and PGF 1.01
\ifx\dpiclw\undefined\newdimen\dpiclw\fi
\global\def\dpicdraw{\draw[line width=\dpiclw]}
\global\def\dpicstop{;}
\dpiclw=0.8bp
\dpicdraw[fill=lightgray](0,-0.15625) rectangle (0.3125,0.15625)\dpicstop
\draw (0.15625,0) node{$0/1$};
\dpicdraw[fill=lightgray](-0.3125,-0.78125) rectangle (0,-0.46875)\dpicstop
\draw (-0.15625,-0.625) node{$a_1$};
\dpicdraw[fill=lightgray](0.3125,-0.78125) rectangle (0.625,-0.46875)\dpicstop
\draw (0.46875,-0.625) node{$a_2$};
\dpicdraw (-0.15625,-0.46875)
 --(-0.15625,-0.3125)
 --(0.46875,-0.3125)
 --(0.46875,-0.46875)\dpicstop
\dpicdraw (0.15625,-0.15625)
 --(0.15625,-0.3125)\dpicstop
\draw (0,0) node[left=-0.46875bp]{Alice};
\dpicdraw[fill=lightgray](1.5625,-0.15625) rectangle (1.875,0.15625)\dpicstop
\draw (1.71875,0) node{$0/1$};
\dpicdraw[fill=lightgray](1.25,-0.78125) rectangle (1.5625,-0.46875)\dpicstop
\draw (1.40625,-0.625) node{$b_1$};
\dpicdraw[fill=lightgray](1.875,-0.78125) rectangle (2.1875,-0.46875)\dpicstop
\draw (2.03125,-0.625) node{$b_2$};
\dpicdraw (1.40625,-0.46875)
 --(1.40625,-0.3125)
 --(2.03125,-0.3125)
 --(2.03125,-0.46875)\dpicstop
\dpicdraw (1.71875,-0.15625)
 --(1.71875,-0.3125)\dpicstop
\draw (1.875,0) node[right=-0.46875bp]{Bob};
\draw (0.9375,1.59375) node[above=-0.46875bp]{Target};
\dpicdraw[dashed](0.15625,0.15625)
 --(0.885348,1.485604)\dpicstop
\filldraw[line width=0bp](0.912747,1.470576)
 --(0.915403,1.540403)
 --(0.857948,1.500631) --cycle
\dpicstop
\draw (0.534893,0.846625) node[left=-0.46875bp]{$a_2 = 1$};
\dpicdraw[dashed](1.71875,0.15625)
 --(0.989652,1.485604)\dpicstop
\filldraw[line width=0bp](1.017052,1.500631)
 --(0.959597,1.540403)
 --(0.962253,1.470576) --cycle
\dpicstop
\draw (1.340107,0.846625) node[right=-0.46875bp]{$b_1 = 0$};
\end{tikzpicture}
\end{center}
\caption{Alice-Bob scenario}
\end{figure}

We now suppose that Alice and Bob perform repeated rounds of these operations. On different rounds, they may make different choices of which measurement settings to use, and they may observe different outcomes for a given choice of setting.
The target can compile statistics for this series of data, and infer probability distributions on the outcomes.
The probability table in Figure~3 records the result of such a process.

\begin{figure}
\label{probtab}
\begin{center}
\begin{tabular}{ll|ccccc}
A & B & $(0, 0)$ & $(1, 0)$ & $(0, 1)$ & $(1, 1)$  &  \\ \hline
$a_1$ & $b_1$ & $1/2$ & $0$ & $0$ & $1/2$ & \\
$a_1$ & $b_2$ & $3/8$ & $1/8$ & $1/8$ & $3/8$ & \\
$a_2$ & $b_1$ & $3/8$ & $1/8$ & $1/8$ & $3/8$ &  \\
$a_2$ & $b_2$ & $1/8$ & $3/8$ & $3/8$ & $1/8$ & 
\end{tabular}
\end{center}
\caption{The Bell table}
\end{figure}

Consider for example the cell at row 2, column 3 of the table. This corresponds to the following event:
\begin{itemize}
\item Alice chooses measurement setting $a_1$ and observes the outcome $0$.
\item Bob chooses measurement setting $b_2$ and observes the outcome $1$.
\end{itemize}
This event has the probability $1/8$, conditioned on Alice's choice of $a_1$ and Bob's choice of $b_2$.

Each row of the table specifies a probability distribution on the possible joint outcomes, conditioned on the indicated choice of settings by Alice and Bob.

We can now ask:

\vsa
\begin{center}
\fbox{How can such an observational scenario be realised?}
\end{center}
\vsa

The obvious classical mechanism we can propose to explain these observations is depicted in Figure~4.

\begin{figure}
%\vspace{0.5in}
\label{sbitreg}
\begin{center}
\begin{tikzpicture}[scale=2.54]
% dpic version 2011.03.17 option -g for TikZ and PGF 1.01
\ifx\dpiclw\undefined\newdimen\dpiclw\fi
\global\def\dpicdraw{\draw[line width=\dpiclw]}
\global\def\dpicstop{;}
\dpiclw=0.8bp
\dpicdraw[fill=lightgray](0,-0.094961) rectangle (0.227906,0.094961)\dpicstop
\draw (0.113953,0) node{$0/1$};
\dpicdraw[fill=lightgray](-0.17093,-0.474804) rectangle (0.018992,-0.284883)\dpicstop
\draw (-0.075969,-0.379844) node{$a_1$};
\dpicdraw[fill=lightgray](0.208914,-0.474804) rectangle (0.398836,-0.284883)\dpicstop
\draw (0.303875,-0.379844) node{$a_2$};
\dpicdraw (-0.075969,-0.284883)
 --(-0.075969,-0.189922)
 --(0.303875,-0.189922)
 --(0.303875,-0.284883)\dpicstop
\dpicdraw (0.113953,-0.094961)
 --(0.113953,-0.189922)\dpicstop
\draw (0,0) node[left=-0.284883bp]{Alice};
\dpicdraw[fill=lightgray](0.987593,-0.094961) rectangle (1.215499,0.094961)\dpicstop
\draw (1.101546,0) node{$0/1$};
\dpicdraw[fill=lightgray](0.816664,-0.474804) rectangle (1.006585,-0.284883)\dpicstop
\draw (0.911625,-0.379844) node{$b_1$};
\dpicdraw[fill=lightgray](1.196507,-0.474804) rectangle (1.386429,-0.284883)\dpicstop
\draw (1.291468,-0.379844) node{$b_2$};
\dpicdraw (0.911625,-0.284883)
 --(0.911625,-0.189922)
 --(1.291468,-0.189922)
 --(1.291468,-0.284883)\dpicstop
\dpicdraw (1.101546,-0.094961)
 --(1.101546,-0.189922)\dpicstop
\draw (1.215499,0) node[right=-0.284883bp]{Bob};
\draw (0.60775,0.968601) node[above=-0.284883bp]{Target};
\dpicdraw[dashed](0.113953,0.094961)
 --(0.575484,0.903194)\dpicstop
\filldraw[line width=0bp](0.591977,0.893776)
 --(0.59432,0.936179)
 --(0.558992,0.912612) --cycle
\dpicstop
\draw (0.353552,0.514546) node[left=-0.284883bp]{$a_2 = 1$};
\dpicdraw[dashed](1.101546,0.094961)
 --(0.640015,0.903194)\dpicstop
\filldraw[line width=0bp](0.656508,0.912612)
 --(0.621179,0.936179)
 --(0.623522,0.893776) --cycle
\dpicstop
\draw (0.861948,0.514546) node[right=-0.284883bp]{$b_1 = 0$};
\dpicdraw[fill=lightgray](0.417828,-1.538366) rectangle (0.512789,-1.348445)\dpicstop
\draw (0.465308,-1.443405) node{$0$};
\dpicdraw[fill=lightgray](0.512789,-1.538366) rectangle (0.60775,-1.348445)\dpicstop
\draw (0.560269,-1.443405) node{$1$};
\dpicdraw[fill=lightgray](0.60775,-1.538366) rectangle (0.702711,-1.348445)\dpicstop
\draw (0.65523,-1.443405) node{$0$};
\dpicdraw[fill=lightgray](0.702711,-1.538366) rectangle (0.797671,-1.348445)\dpicstop
\draw (0.750191,-1.443405) node{$1$};
\dpicdraw[fill=lightgray](0.417828,-1.91821) rectangle (0.797671,-1.538366)\dpicstop
\draw (0.60775,-1.728288) node{$\vdots$};
\dpicdraw[dashed](0.465308,-1.348445)
 --(-0.055963,-0.507094)\dpicstop
\filldraw[line width=0bp](-0.039819,-0.497091)
 --(-0.075969,-0.474804)
 --(-0.072108,-0.517096) --cycle
\dpicstop
\dpicdraw[dashed](0.560269,-1.348445)
 --(0.314571,-0.511252)\dpicstop
\filldraw[line width=0bp](0.332795,-0.505903)
 --(0.303875,-0.474804)
 --(0.296348,-0.5166) --cycle
\dpicstop
\dpicdraw[dashed](0.65523,-1.348445)
 --(0.900928,-0.511252)\dpicstop
\filldraw[line width=0bp](0.919152,-0.5166)
 --(0.911625,-0.474804)
 --(0.882704,-0.505903) --cycle
\dpicstop
\dpicdraw[dashed](0.750191,-1.348445)
 --(1.271463,-0.507094)\dpicstop
\filldraw[line width=0bp](1.287607,-0.517096)
 --(1.291468,-0.474804)
 --(1.255318,-0.497091) --cycle
\dpicstop
\draw (0.60775,-1.91821) node[below=-0.284883bp]{Source};
\end{tikzpicture}
\end{center}
\caption{A Source}
\end{figure}
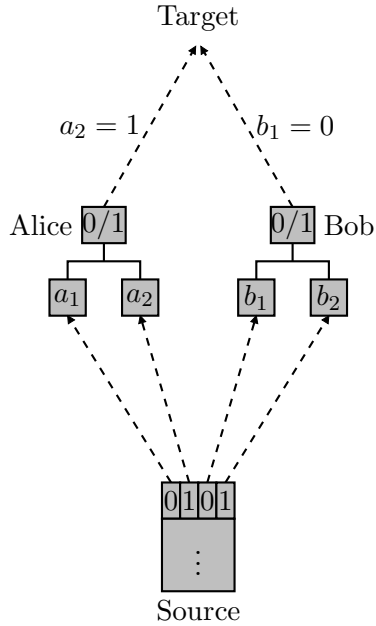
We postulate a \emph{source} which on each round chooses outcomes for each of the possible measurement settings $a_1$, $a_2$, $b_1$, $b_2$. Alice and Bob will then observe the values which have been chosen by the source.
We can suppose that this source is itself randomised, and chooses the outcomes according to some probability distribution $P$ on the set of $2^{4}$ possible assignments.

We can now ask the question: is there any distribution $P$ which would give rise to the table specified in Figure~2?

\paragraph{Important Note}
A key observation is that, in order for this question to be non-trivial, we must assume that the choices of measurement settings made by Alice and Bob are \emph{independent} of the source.\footnote{This translates formally into a conditional independence assumption, which we shall not spell out here; see e.g. \citep{Bellbeables}, \citep{brandenburger2008classification}.} If the source  could determine which measurements are to be performed on each round, as well as their values, then it becomes a trivial matter to achieve \emph{any} given probability distribution on the joint outcomes.

Under this assumption of independence, it becomes natural to think of this scenario as a kind of \emph{correlation game}. The aim of the source is to achieve as high a degree of correlation between the outcomes of Alice and Bob as possible, whatever the choices made by Alice and Bob on each round.

\subsection{Logic rings a Bell}
We shall now make a very elementary and apparently innocuous deduction in elementary logic and probability theory, which could easily be carried out by students in the first few weeks of a Probability~$101$ course.

Suppose we have propositional formulas $\vphi_1, \ldots , \vphi_N$.
We suppose further that we can assign a probability $p_i$ to each $\vphi_i$.

In particular, we have in the mind the situation where the boolean variables appearing in $\vphi_i$ correspond to empirically testable quantities, such as the outcomes of measurements in our scenario; $\vphi_i$ then expresses a condition on the outcomes of an
experiment involving these quantities.
The probabilities $p_i$ are obtained from the statistics of these experiments.

Now suppose that these formulas are \emph{not simultaneously satisfiable}.  Then (e.g.)
\[ \bigwedge_{i=1}^{N-1} \phi_i \IMP \neg \phi_N ,  \quad \mbox{or equivalently} 
\quad \phi_N \IMP \bigvee_{i=1}^{N-1} \neg \phi_i . \]
Using elementary probability theory, we can calculate:
 \[ p_N  \; \leq \; \prob(\bigvee_{i=1}^{N-1} \neg \phi_i) \; \leq \; \sum_{i=1}^{N-1} \prob(\neg \phi_i) \; = \; \sum_{i=1}^{N-1} (1 - p_i) \; = \; (N - 1) - \sum_{i=1}^{N-1} p_i . \]
The first inequality is the monotonicity of probability, and the second is sub-additivity.

Hence we obtain the inequality
 \[ \sum_{i=1}^N p_i \; \leq \; N-1. \]
We shall refer to this as a \emph{logical Bell inequality}, for reasons to be discussed later. Note that it hinges on a purely logical consistency condition.

\subsection{Logical analysis of the Bell table}

We return to the probability table from Figure~2.

\begin{center}
\begin{tabular}{l|ccccc}
& $(0, 0)$ & $(1, 0)$ & $(0, 1)$ & $(1, 1)$  &  \\ \hline
$(a_1, b_1)$ & { \fcolorbox{gray}{lightgray}{1/2}} & $0$ & $0$ & { \fcolorbox{gray}{lightgray}{1/2}} & \\
$(a_1, b_2)$ &  { \fcolorbox{gray}{lightgray}{3/8}} & $1/8$ & $1/8$ & { \fcolorbox{gray}{lightgray}{3/8}} & \\
$(a_2, b_1)$ & { \fcolorbox{gray}{lightgray}{3/8}} & $1/8$ & $1/8$ & { \fcolorbox{gray}{lightgray}{3/8}} &  \\
$(a_2, b_2)$ & $1/8$ & { \fcolorbox{gray}{lightgray}{3/8}} & { \fcolorbox{gray}{lightgray}{3/8}} & $1/8$ & 
\end{tabular}
\end{center}

If we read $0$ as true and $1$ as false, the highlighted entries in each row of the table are represented by the following propositions:
\[ \begin{array}{rcccccccc}
\vphi_1 & = &  (a_{1} \wedge b_{1}) & \vee & (\neg a_{1} \wedge \neg b_{1}) & = & a_{1} & \leftrightarrow & b_{1} \\
\vphi_2 & = & (a_{1} \wedge b_{2}) & \vee & (\neg a_{1} \wedge \neg b_{2}) & = &  a_{1} & \leftrightarrow & b_{2} \\
\vphi_3 & = & (a_{2} \wedge b_{1}) & \vee & (\neg a_{2} \wedge \neg b_{1}) & = &  a_{2} & \leftrightarrow & b_{1} \\
\vphi_4 & = & (\neg a_{2} \wedge b_{2}) & \vee & (a_{2} \wedge \neg b_{2}) & = & a_{2} & \oplus & b_{2} .
\end{array}
\]
The events on first three rows are the correlated outcomes; the fourth is anticorrelated.
These propositions are easily seen to be jointly unsatisfiable. Indeed, starting with
$\vphi_4$, we can replace $a_{2}$ with $b_{1}$ using $\vphi_3$, $b_{1}$ with $a_{1}$ using $\vphi_1$, and $a_{1}$ with $b_{2}$ using $\vphi_2$, to obtain $b_{2} \oplus b_{2}$, which is obviously unsatisfiable.

It follows that our logical Bell inequality should apply, yielding the inequality
\[ \sum_{i=1}^4 p_i \; \leq \; 3. \]
However, we see from the table that $p_1 = 1$, $p_i = 6/8$ for $i = 2, 3, 4$.
Hence the table yields a violation of the Bell inequality by $1/4$.

This rules out the possibility of giving an explanation for the observational behaviour described by the table in terms of a classical source.
We might then conclude that such behaviour simply cannot be realised. However,  \emph{in the presence of quantum resources, this is no longer the case}.
More specifically, if we use the Bell state  $(\ket{\ua\ua} + \ket{\da\da})/\SQT$, with Alice and Bob performing 1-qubit local measurements corresponding to directions in the $XY$-plane of the Bloch sphere, at relative angle $\nicefrac{\pi}{3}$,
then this behaviour is \emph{physically realisable}  according to the predictions of quantum mechanics---%
our most highly-confirmed physical theory.\footnote{For further details on this, see e.g.~\citep{Abr15}.}

More broadly, we can say that this shows that quantum mechanics predicts correlations which exceed those which can be achieved by any classical mechanism. This is the content of \emph{Bell's theorem} \citep{bell:64}, a famous result in the foundations of quantum mechanics, and in many ways the starting point for the whole field of quantum information.
Moreover, these predictions have been confirmed by many experiments \citep{aspect:82}, \citep{aspect:99}.

The logical Bell inequality we used to derive this result is taken from \citep{abramsky:12}. Bell inequalities are a central technique in quantum information and foundations. In \citep{abramsky:12}  it is shown that \emph{every} Bell inequality (\ie every inequality satisfied by the ``local polytope'') is equivalent to a logical Bell inequality, based on purely logical consistency conditions.

\section{More on formalization}

We note that the example in the previous section, which is representative of those studied in quantum contextuality and non-locality, has a specific form which can be reflected naturally in our mathematical formalism.

In particular, the ``features'' take the form of a set $X$ of variables, which can be measured or observed. Thus we take $\Feat = X$. The set $\Cont$ consists of compatible sets of variables, those which can be measured jointly. The result of measuring a compatible set of variables $C$ is a \emph{joint outcome} in $\prod_{x \in C} O_x$, where $O_x$ is the set of possible outcomes or values for the variable $x \in X$. To simplify notation, we shall assume there is a single set of outcomes $O$ for all variables, so a joint outcome for $C$ is an element of $O^C$ --- an assignment of a value in $O$ to each $x \in C$. By repeatedly observing joint outcomes for the variables in a context $C$, we obtain statistics, from which we can infer a probability distribution $d \in \prob(O^C)$, where $\prob(X)$ is the set of probability distributions on a set $X$.\footnote{To avoid measure-theoretic technicalities, we shall assume that we are dealing with discrete distributions. In fact, in all the examples we shall consider, the sets $X$ and $O$ will be finite.} So the presheaf $P$ has the specific form $P(C) = \prob(O^C)$.
What about the restriction maps?

Note firstly that the assignment $C \mapsto O^C$ has a natural contravariant functorial action (it is a restriction of the contravariant hom functor). If $C \subseteq D$, then restriction  is just function restriction to a subset of the domain:
\[ \rho^D_C : O^D \rTo O^C, \qquad \rho^D_C(f) = f |_C . \]
Moreover, the assignment $X \mapsto \prob(X)$ can be extended to a (covariant) functor $\DD : \Set \rTo \Set$:
$\DD(X) = \prob(X)$, and given $f : X \rTo Y$, $\DD(f) : \DD(X) \rTo \DD(Y)$ is the push-forward of probability measures along $f$:
\[ \DD(f)(d)(U) \; = \; d(f^{-1}(U)) . \]
This functor extends to a monad on $\Set$; the discrete version of the Giry monad \citep{giry1982categorical}.

Composing these two functors, we can define a presheaf $P : \pow(X)^{\op} \rTo \Set$, with $P(C) = \DD(O^C)$, and if $C \subseteq D$, $d \in \DD(O^D)$:
\[ \rho^D_C(d)(U) = d(\{ s \in O^D \mid s |_C \in U \}) . \]
Note that if $C \subseteq D$, we can write $D$ as a disjoint union $D = C \sqcup C'$, and $O^D = O^C \times O^{C'}$. Restriction of $s \in O^D$ to $C$ is projection onto the first factor of this product. Thus restriction of a distribution is \emph{marginalization}.

A compatible family of data for this presheaf is a family $\{ d_C \}_{C \in \Cont}$ of probability distributions, $d_C \in \DD(O^C)$. Local consistency, \ie compatibility of the family, is the condition that for all $C, D \in \Cont$,
\[ d_C |_{C \cap D} \; = \; d_D |_{C \cap D} . \]
This says that the distributions on $O^C$ and $O^D$ have the same marginals on their overlap, \ie the common factor $O^{C \cap D}$.

Global consistency is exactly the condition that there is a \emph{joint distribution} $d \in \DD(O^X)$ from which the distributions $d_C$ can be recovered by marginalization.

We refer again to the Alice-Bob table in Figure~3. We anatomize this table according to our formal scheme. The set of variables is $X = \{ a_1, a_2, b_1, b_2 \}$.
The set of contexts, indexing the rows of the table, is 
\[ \Cont \; = \; \{ \{ a_1, b_1 \}, \quad \{ a_2, b_1 \}, \quad \{ a_1, b_2 \}, \quad \{ a_2, b_2 \} \} .\]
These are the compatible sets of measurements. The set of outcomes is $O = \{ 0, 1 \}$.
The columns are indexed by the possible joint outcomes of the measurements. Thus for example the matrix entry at row $(a_2, b_1)$ and column $(0,1)$ 
indicates the  \emph{event }
\[ \{ a_2 \mapsto 0,  \; b_1 \mapsto 1 \} . \]
The set of events relative to a context $C$ is the set of functions $O^C$.
Each row of the table, indexed by context $C$, specifies a distribution $d_C \in \DD(O^C)$. Thus the set of rows of the table is the family of data $\{ d_C \}_{C \in \Cont}$.
One can check that this family is locally consistent, \ie compatible. In this example, this is exactly the No-Signalling principle \citep{ghirardi:80,popescu:94}, a fundamental physical principle which is satisfied by quantum mechanics. Suppose that $C = \{ a, b \}$, and $C' = \{ a, b' \}$, where $a$ is a variable measured by Alice, while $b$ and $b'$ are variables measured by Bob. Then under relativistic constraints, Bob's choice of measurement --- $b$ or $b'$ --- should not be able to affect the distribution Alice observes on the outcomes from her measurement of $a$. This is captured by the No-Signalling principle, which says that the distribution on $\{ a \} = \{ a, b \} \cap \{ a, b' \}$ is the same whether we marginalize from the  distribution  $e_C$, or the distribution $e_{C'}$. This is exactly compatibility.

Thus we can see that this table, and the others like it which are standardly studied in non-locality and contextuality theory within quantum information and foundations, fall exactly within the scope of our definition.

At the same time, we can regard \emph{any} instance of our general definition of a family of data as a generalized probability table; compatibility becomes a generalized form of No-Signalling (which, suitably formulated, can be shown to hold in this generality in quantum mechanics \citep{abramsky:11}); and we have a very general probabilistic concept of contextuality.

It might appear from our discussion thus far that contextuality is inherently linked to probabilistic behaviour. However, this is not the case. In fact, our initial formalization in Section~2 was considerably more general; and even within the current more specialized version, we can gain a much wider perspective by  generalizing the notion of distribution.
We recall firstly that a probability distribution of finite support on a  set $X$ can be specified as a function
\[ d : X \rTo \Rgeq \]
where $\Rgeq$ is the set of non-negative reals, 
satisfying the normalization condition
\[ \sum_{x \in X} d(x) \; = \; 1 . \]
This condition guarantees that the range of the function lies within the unit interval $[0, 1]$. The finite support condition means that $d$ is zero on all but a finite subset  of $X$. The probability assigned to an event $E \subseteq X$ is then given by
\[ d(E) \; = \; \sum_{x \in E} d(x) . \]

This is easily generalized by replacing $\Rgeq$ by an arbitrary  \emph{commutative semiring}, which is an algebraic structure $(R, {+}, 0, {\cdot}, 1)$, where $(R, {+}, 0)$ and $(R, {\cdot}, 1)$ are commutative monoids satisfying the distributive law:
\[ a \cdot (b + c) = a\cdot b + a \cdot c . \]
Examples include the non-negative reals $\Rgeq$ with the usual addition and multiplication, and the booleans $\Bool = \{ 0, 1 \}$ with disjunction and conjunction playing the r\^oles of addition and multiplication respectively.

We can now define a functor $\DD_R$ of $R$-distributions, parameterized by a commutative semiring $R$. Given a set $X$, $\DD_R(X)$ is the set of $R$-distributions of finite support. The functorial action is defined exactly as we did for $\DD = \DD_{\Rgeq}$.
In the boolean case, $\Bool$-distributions on $X$ correspond to non-empty finite subsets of $X$. In this boolean case, we have a notion of \emph{possibilistic contextuality}, where we have replaced probabilities by boolean values, corresponding to possible or impossible.
Note that there is a homomorphism of semirings from $\Rgeq$ to $\Bool$, which sends positive probabilities to $1$ (possible), and $0$ to $0$ (impossible). This lifts to a map on distributions, which sends a probability distribution to its \emph{support}. This in turn sends generalized probability tables $\{ d_C \}_{C \in \Cont}$ to possibility tables.
We refer to this induced map as the \emph{possibilistic collapse}.

\subsection{Measurement Scenarios and Empirical Models}
To understand how contextuality behaves across this possibilistic collapse, it will be useful to introduce some terminology.
The combinatorial shape of the situations we are considering is determined by the following data:
\begin{itemize}
\item The set $X$ of variables
\item The set $\Cont$ of contexts, \ie the compatible sets of variables
\item The set of outcomes $O$.
\end{itemize}
Accordingly, we shall call $(X, \Cont, O)$ a \emph{measurement scenario}.
Once we have fixed a measurement scenario $\Sigma = (X, \Cont, O)$, and a semiring $R$, we have the notion of a compatible family of $R$-distributions $\{ e_C \}_{C \in \Cont}$, where $e_C \in \DD_R(O^C)$. We will call such compatible families \emph{empirical models}, and write $\EM(\Sigma, R)$ for the set of empirical models over the scenario $\Sigma$ and the semiring $R$. We refer to \emph{probabilistic empirical models} for $R = \Rgeq$, and \emph{possibilistic empirical models} for $R = \Bool$.

Given an empirical model $e \in \EM(\Sigma, R)$, we call a distribution $d_g \in \DD_R(O^X)$ such that, for all $C \in \Cont$, $d_g |_C = e_C$, a \emph{global section} for $e$.
The existence of such a global section is precisely our formulation of global consistency.
Thus 
\begin{center}
\vsa
\fbox{\emph{$e$ is contextual if and only if it has no global section}}
\vsa
\end{center}

The term ``empirical model'' relates to the fact that this is purely observational data, without prejudice as to what mechanism or physical theory lies behind it.
In the terminology of quantum information, it is a \emph{device-independent} notion \citep{barrett2005no}.
It is a key point that \emph{contextuality is in the data}. It does not presuppose quantum mechanics, or any other specific theory.

A homomorphism of semirings $h : R \rTo S$ induces a natural transformation $\hb$ from the presheaf of $R$-valued distributions to the presheaf of $S$-valued distributions.
In particular, if $h : \Rgeq \rTo \Bool$ is the unique semiring homomorphism from the positive reals to the booleans, then $\hb$ is the possibilistic collapse.
Given a global section $d_g$ for an empirical model $e \in EM(\Sigma, R)$, it is easy to see that $\hb(d_g)$ is a global section for $\hb(e)$.
Thus we have the following result.

\begin{proposition}
\label{posscontprop}
If $\hb(e)$ is contextual, then so is $e$. In particular, if the possibilistic collapse of a probabilistic empirical model $e$ is contextual, then $e$ is contextual.
\end{proposition}

This is a strict one-way implication. It is strictly harder to be possibilistically contextual than probabilistically contextual, as we shall now see.

\section{Example: The Hardy Paradox}

Consider the table in Figure~5.

\begin{figure}
%\vspace{0.75in}
\label{Hardy}
\begin{center}
\begin{tabular}{c|cccc} 
 & $(0, 0)$ & $(0,1)$ & $(1,0)$ & $(1, 1)$ \\ \hline
$(a_1, b_1)$ &  $1$ &  &  &  \\
$(a_1, b_2)$ &   $0$ &  &  &  \\
$(a_2, b_1)$ &  $0$ &  &  & \\
$(a_2, b_2)$ &   &  &  & $0$ \\
\end{tabular}
\end{center}
\caption{The Hardy Paradox}
\end{figure}

This table depicts the same kind of scenario we considered previously. However, the entries are now either $0$ or $1$. The idea is that a $1$ entry represents a positive probability. Thus we are distinguishing only between \emph{possible} (positive probability) and \emph{impossible} (zero probability). In other words, the rows correspond to the \emph{supports} of some (otherwise unspecified) probability distributions.
Moreover, only four entries of the table are filled in. Our claim is that just from these four entries, referring only to the supports, we can deduce that  the behaviour recorded in this table is contextual. Moreover, this behaviour can again be realised in quantum mechanics, yielding a stronger form of Bell's theorem, due to Lucien Hardy \citep{hardy:93}.\footnote{For a detailed discussion of  realisations of the Bell and Hardy models in quantum mechanics, see Section~7 of \citep{abramsky2013rhv}. Further details on the Hardy construction can be found in a number of papers\citep{hardy:93,mermin1994quantum}.}

\subsection{What Do ``Observables'' Observe?}

Classically, we would take the view that physical observables directly reflect properties of the physical system we are observing. These are objective properties of the system, which are independent of our choice of which measurements to perform ---
of our \emph{measurement context}.
More precisely, this would say that for each possible state of the system, there is a function $\lambda$ which for each measurement $m$ specifies an outcome $\lambda(m)$, \emph{independently of which other measurements may be performed}.
This point of view is called \emph{non-contextuality}, and may seem self-evident.
However, this view is \emph{impossible to sustain} in the light of our \emph{actual observations of (micro)-physical reality}.

Consider once again the Hardy table depicted in Figure~5. Suppose there is a function $\lambda$ which accounts for the possibility of Alice observing value $0$ for $a_1$ and Bob observing $0$ for $b_1$, as asserted by the entry in the top left position in the table. Then this function $\lambda$ must satisfy
 \[ \lambda : a_1 \mapsto 0, \quad b_1 \mapsto 0 . \]
Now consider the value of $\lambda$ at $b_2$. If $\lambda(b_2) = 0$, then this would imply that the event that $a_1$ has value $0$ and $b_2$ has value $0$ is possible. However, \emph{this is precluded} by the $0$ entry in the table for this event. The only other possibility is that $\lambda(b_2) = 1$. Reasoning similarly with respect to the joint values of $a_2$ and $b_2$, we conclude, using the bottom right entry in the table, that we must have $\lambda(a_2) = 0$. Thus the only possibility for $\lambda$ consistent with these entries is
\[ \lambda : a_1 \mapsto 0, \quad a_2 \mapsto 0, \quad b_1 \mapsto 0, \quad b_2 \mapsto 1 . \]
However, this would require the outcome $(0, 0)$ for measurements $(a_2,b_1)$ to be possible, and this is \emph{precluded} by the table.

We are thus forced to conclude that the Hardy models are contextual. Moreover,  it is  \emph{possibilistically contextual}. By virtue of Proposition~\ref{posscontprop}, we know that any probabilstic model with this support must be probabilistically contextual. On the other hand, if we consider the support of the table in Figure~3, it is easy to see that it is not possibilistically contextual. Thus we see that possibilistic contextuality is strictly stronger than probabilistic contextuality.

\section{Strong Contextuality}

Logical contextuality as exhibited by the Hardy paradox can be expressed in the following form: there is a local assignment (in the Hardy case, the assignment $a_1 \mapsto 0, b_1 \mapsto 0$) which is in the support, but which cannot be extended to a global assignment which is compatible with the support. This says that the support cannot be covered by the projections of global assignments. 
A stronger form of contextuality is when \emph{no global assignments are consistent with the support at all}.
Note that this stronger form does not hold for the Hardy paradox.

We pause to state this in more precise terms. Consider a possibilistic model  $e \in \EM(\Sigma, \Bool)$. For example, $e$ may be the possibilistic collapse of a probabilistic model.
We can consider the boolean distribution $e_C$ as (the characteristic function of) a subset of $O^C$. In fact, the compatibility of $e$ implies that it is a sub-presheaf of $C \mapsto O^C$. We write $S_e(C) \subseteq O^C$ for this subpresheaf.
A global assignment $g \in O^X$ is \emph{consistent with $e$} if $g |_C \in S_e(C)$ for all $C \in \Cont$. We write $S_e(X)$ for the set of such global assignments.
A \emph{compatible family of local assignments for $e$} is a family $\{ s_C \}_{C \in \Cont}$ with $s_C \in S_e(C)$, such that for all $C, D \in \Cont$, $s_C |_{C \cap D} = s_D |_{C \cap D}$.

\begin{proposition}
There is a bijective correspondence between $S_e(X)$ and compatible families of local assignments for $e$.
\end{proposition}
This follows directly from the observation that the presheaf $C \mapsto O^C$ is a sheaf, and hence $S_e$ as a subpresheaf is separated.

We can now characterize possibilistic and strong contextuality in terms of the extendability of local assignments to global ones.
\begin{proposition}
Let $e \in \EM(\Sigma, \Bool)$ be a possibilistic empirical model.
The following are equivalent:
\begin{enumerate}
\item $e$ is possibilistically contextual.
\item There exists a local assignment $s \in S_e(C)$ which cannot be extended to a compatible family of local assignments.
\item There exists a local assignment $s \in S_e(C)$ such that, for all $g \in S_e(X)$, $g |_C \neq s$.
\end{enumerate}
Also, the following are equivalent:
\begin{enumerate}
\item $e$ is strongly contextual.
\item No local assignment $s \in S_e(C)$ can be extended to a compatible family of local assignments.
\item $S_e(X) = \vn$.
\end{enumerate}
\end{proposition}

Several much-studied constructions from the quantum information literature exemplify strong contextuality.
An important example is the Popescu--Rohrlich (PR) box \citep{popescu:94} shown in Fig.~\ref{fig:PR}.

\begin{figure}
\begin{center}
\small
\begin{tabular}{ll|ccccc}
A & B & $(0, 0)$ & $(1, 0)$ & $(0, 1)$ & $(1, 1)$  &  \\ \hline
$a_1$ & $b_1$ & $1$ & $0$ & $0$ & $1$ & \\
$a_1$ & $b_2$ & $1$ & $0$ & $0$ & $1$ & \\
$a_2$ & $b_1$ & $1$ & $0$ & $0$ & $1$ & \\
$a_2$ & $b_2$ & $0$ & $1$ & $1$ & $0$ & 
\end{tabular}
\end{center}
\caption{The PR Box}
\label{fig:PR}
\end{figure}%
This is a behaviour which satisfies the \emph{no-signalling principle} \citep{ghirardi:80}, meaning that the probability of Alice observing a particular outcome for her choice of measurement (e.g. $a_1=0$), is independent of whether Bob chooses measurement $b_1$ or $b_2$; and vice versa. 
That is, Alice and Bob cannot signal to one another---the importance of this principle is that it enforces compatibility with relativistic constraints.
However, despite satisfying the no-signalling principle, the PR box does not admit a quantum realisation. Note that the full support of this model correspond to the propositions used in showing the contextuality of the Bell table from Fig.~3, and hence the fact that these propositions are not simultaneously satisfiable shows the strong contextuality of the model.

In fact, there is provably no bipartite quantum-realisable behaviour of this kind which is strongly contextual \citep{lal:11,mansfield:14}. 
However, as soon as we go to three or more parties, strong contextuality does arise from entangled quantum states. A notable example is provided by the GHZ states \citep{greenberger:90}.
Thus we have a strict hierarchy of strengths of contextuality, 
\[ \mbox{probabilistic} \quad < \quad \mbox{possibilistic} \quad < \quad \mbox{strong} \]
exemplified in terms of well-known examples from the quantum foundations literature as follows:
\[ \mbox{Bell} \quad < \quad \mbox{Hardy} \quad < \quad \mbox{GHZ} . \]

\section{Visualizing Contextuality}

The tables which have appeared in our examples can be displayed in a visually appealing way which makes the fibred topological structure apparent, and forms an intuitive bridge to the formal development of the sheaf-theoretic ideas.

\begin{figure}
\begin{center}
\begin{tikzpicture}[x=50pt,y=50pt,thick,label distance=-0.25em,baseline=(O.base)]
\coordinate (O) at (0,0);
\coordinate (T) at (0,1.5);
\coordinate (u) at (0,0.5);
\coordinate [inner sep=0em] (v0) at ($ ({-cos(1*pi/12 r)*1.2},{-sin(1*pi/12 r)*0.48}) $);
\coordinate [inner sep=0em] (v1) at ($ ({-cos(7*pi/12 r)*1.2},{-sin(7*pi/12 r)*0.48}) $);
\coordinate [inner sep=0em] (v2) at ($ ({-cos(13*pi/12 r)*1.2},{-sin(13*pi/12 r)*0.48}) $);
\coordinate [inner sep=0em] (v3) at ($ ({-cos(19*pi/12 r)*1.2},{-sin(19*pi/12 r)*0.48}) $);
\coordinate [inner sep=0em] (v0-1) at ($ (v0) + (T) $);
\coordinate [inner sep=0em] (v0-0) at ($ (v0-1) + (u) $);
\coordinate [inner sep=0em] (v1-1) at ($ (v1) + (T) $);
\coordinate [inner sep=0em] (v1-0) at ($ (v1-1) + (u) $);
\coordinate [inner sep=0em] (v2-1) at ($ (v2) + (T) $);
\coordinate [inner sep=0em] (v2-0) at ($ (v2-1) + (u) $);
\coordinate [inner sep=0em] (v3-1) at ($ (v3) + (T) $);
\coordinate [inner sep=0em] (v3-0) at ($ (v3-1) + (u) $);
\draw (v0) -- (v1) -- (v2) -- (v3) -- (v0);
\draw [dotted] (v0-0) -- (v0);
\draw [dotted] (v1-0) -- (v1);
\draw [dotted] (v2-0) -- (v2);
\draw [dotted] (v3-0) -- (v3);
\node [inner sep=0.1em] (v0') at (v0) {$\bullet$};
\node [anchor=east,inner sep=0em] at (v0'.west) {$a_1$};
\node [inner sep=0.1em,label={[label distance=-0.625em]330:{$b_1$}}] at (v1) {$\bullet$};
\node [inner sep=0.1em] (v2') at (v2) {$\bullet$};
\node [anchor=west,inner sep=0em] at (v2'.east) {$a_2$};
\node [inner sep=0.1em,label={[label distance=-0.5em]175:{$b_2$}}] at (v3) {$\bullet$};
\draw [line width=3.2pt,white] (v0-0) -- (v3-1);
\draw [line width=3.2pt,white] (v0-1) -- (v3-0);
\draw [line width=3.2pt,white] (v0-1) -- (v3-1);
\draw (v0-0) -- (v3-1);
\draw [blue] (v0-1) -- (v3-0);
\draw (v0-1) -- (v3-1);
\draw [line width=3.2pt,white] (v2-0) -- (v3-0);
\draw [line width=3.2pt,white] (v2-0) -- (v3-1);
\draw [line width=3.2pt,white] (v2-1) -- (v3-0);
\draw [blue] (v2-0) -- (v3-0);
\draw (v2-0) -- (v3-1);
\draw (v2-1) -- (v3-0);
\draw [line width=3.2pt,white] (v0-0) -- (v1-0);
\draw [line width=3.2pt,white] (v0-0) -- (v1-1);
\draw [line width=3.2pt,white] (v0-1) -- (v1-0);
\draw [line width=3.2pt,white] (v0-1) -- (v1-1);
\draw [red] (v0-0) -- (v1-0);
\draw (v0-0) -- (v1-1);
\draw (v0-1) -- (v1-0);
\draw [blue] (v0-1) -- (v1-1);
\draw [line width=3.2pt,white] (v2-0) -- (v1-1);
\draw [line width=3.2pt,white] (v2-1) -- (v1-0);
\draw [line width=3.2pt,white] (v2-1) -- (v1-1);
\draw [blue] (v2-0) -- (v1-1);
\draw (v2-1) -- (v1-0);
\draw (v2-1) -- (v1-1);
\node [inner sep=0.1em,label=left:{$0$}] at (v0-0) {$\bullet$};
\node [inner sep=0.1em,label=left:{$1$}] at (v0-1) {$\bullet$};
\node [inner sep=0.1em] at (v1-0) {$\bullet$};
\node [inner sep=0.1em,label={[label distance=-0.5em]330:{$1$}}] at (v1-1) {$\bullet$};
\node [inner sep=0.1em,label=right:{$0$}] at (v2-0) {$\bullet$};
\node [inner sep=0.1em,label=right:{$1$}] at (v2-1) {$\bullet$};
\node [inner sep=0.1em,label={[label distance=-0.5em]150:{$0$}}] at (v3-0) {$\bullet$};
\node [inner sep=0.1em] at (v3-1) {$\bullet$};
\end{tikzpicture}
\hfil%
\begin{tikzpicture}[x=50pt,y=50pt,thick,label distance=-0.25em,baseline=(O.base)]
\coordinate (O) at (0,0);
\coordinate (T) at (0,1.5);
\coordinate (u) at (0,0.5);
\coordinate [inner sep=0em] (v0) at ($ ({-cos(1*pi/12 r)*1.2},{-sin(1*pi/12 r)*0.48}) $);
\coordinate [inner sep=0em] (v1) at ($ ({-cos(7*pi/12 r)*1.2},{-sin(7*pi/12 r)*0.48}) $);
\coordinate [inner sep=0em] (v2) at ($ ({-cos(13*pi/12 r)*1.2},{-sin(13*pi/12 r)*0.48}) $);
\coordinate [inner sep=0em] (v3) at ($ ({-cos(19*pi/12 r)*1.2},{-sin(19*pi/12 r)*0.48}) $);
\coordinate [inner sep=0em] (v0-1) at ($ (v0) + (T) $);
\coordinate [inner sep=0em] (v0-0) at ($ (v0-1) + (u) $);
\coordinate [inner sep=0em] (v1-1) at ($ (v1) + (T) $);
\coordinate [inner sep=0em] (v1-0) at ($ (v1-1) + (u) $);
\coordinate [inner sep=0em] (v2-1) at ($ (v2) + (T) $);
\coordinate [inner sep=0em] (v2-0) at ($ (v2-1) + (u) $);
\coordinate [inner sep=0em] (v3-1) at ($ (v3) + (T) $);
\coordinate [inner sep=0em] (v3-0) at ($ (v3-1) + (u) $);
\draw (v0) -- (v1) -- (v2) -- (v3) -- (v0);
\draw [dotted] (v0-0) -- (v0);
\draw [dotted] (v1-0) -- (v1);
\draw [dotted] (v2-0) -- (v2);
\draw [dotted] (v3-0) -- (v3);
\node [inner sep=0.1em] (v0') at (v0) {$\bullet$};
\node [anchor=east,inner sep=0em] at (v0'.west) {$a_1$};
\node [inner sep=0.1em,label={[label distance=-0.625em]330:{$b_1$}}] at (v1) {$\bullet$};
\node [inner sep=0.1em] (v2') at (v2) {$\bullet$};
\node [anchor=west,inner sep=0em] at (v2'.east) {$a_2$};
\node [inner sep=0.1em,label={[label distance=-0.5em]175:{$b_2$}}] at (v3) {$\bullet$};
\draw [line width=3.2pt,white] (v0-0) -- (v3-0);
\draw [line width=3.2pt,white] (v0-1) -- (v3-1);
\draw (v0-0) -- (v3-0);
\draw (v0-1) -- (v3-1);
\draw [line width=3.2pt,white] (v2-0) -- (v3-1);
\draw [line width=3.2pt,white] (v2-1) -- (v3-0);
\draw (v2-0) -- (v3-1);
\draw (v2-1) -- (v3-0);
\draw [line width=3.2pt,white] (v0-0) -- (v1-0);
\draw [line width=3.2pt,white] (v0-1) -- (v1-1);
\draw (v0-0) -- (v1-0);
\draw (v0-1) -- (v1-1);
\draw [line width=3.2pt,white] (v2-0) -- (v1-0);
\draw [line width=3.2pt,white] (v2-1) -- (v1-1);
\draw (v2-0) -- (v1-0);
\draw (v2-1) -- (v1-1);
\node [inner sep=0.1em,label=left:{$0$}] at (v0-0) {$\bullet$};
\node [inner sep=0.1em,label=left:{$1$}] at (v0-1) {$\bullet$};
\node [inner sep=0.1em] at (v1-0) {$\bullet$};
\node [inner sep=0.1em,label={[label distance=-0.5em]330:{$1$}}] at (v1-1) {$\bullet$};
\node [inner sep=0.1em,label=right:{$0$}] at (v2-0) {$\bullet$};
\node [inner sep=0.1em,label=right:{$1$}] at (v2-1) {$\bullet$};
\node [inner sep=0.1em,label={[label distance=-0.5em]150:{$0$}}] at (v3-0) {$\bullet$};
\node [inner sep=0.1em] at (v3-1) {$\bullet$};
\end{tikzpicture}
\end{center}
\caption{The Hardy table and the PR box as bundles}
\label{fig:bundle}
\end{figure}
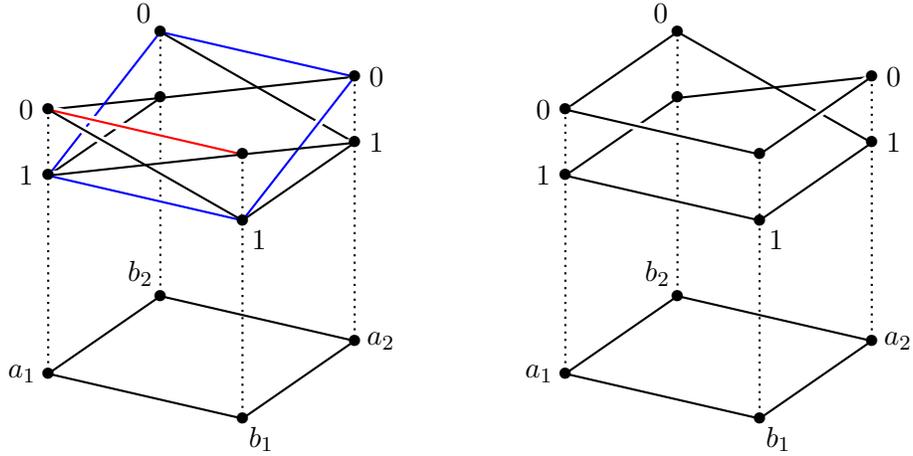

Firstly, we look at the Hardy table from Fig.~5, displayed as a ``bundle diagram'' on the left of Fig.~\ref{fig:bundle}.
Note that all unspecified entries of the Hardy table are set to $1$.

What we see in this representation is the \emph{base space} of the variables $a_1$, $a_2$, $b_1$, $b_2$. There is an edge between two variables when they can be measured together. The pairs of co-measurable variables correspond to the rows of the table. In terms of quantum theory, these correspond to pairs of \emph{compatible observables}. Above each vertex  is a \emph{fibre} of those values which can be assigned to the variable---in this example, $0$ and $1$ in each fibre. There is an edge between values in adjacent fibres precisely when the corresponding \emph{joint outcome} is possible, \ie has a $1$ entry in the table. Thus there are three edges for each of the pairs $\{ a_1, b_2 \}$, $\{ a_2, b_1 \}$ and $\{ a_2, b_2 \}$.

Note that compatibility is expressed topologically by the fact that paths through the fibres can always be extended. That is, if we have an edge over $\{a, b \}$ then there must be an edge over $\{ b, a' \}$ with a common value for $b$, and similarly an edge over $\{ a, b' \}$ with a common value for $a$. This is a local condition.

A \emph{global assignment} corresponds to a closed path traversing all the fibres exactly once. We call such a path \emph{univocal} since it assigns a unique value to each variable. Note that there is such a path, which assigns $1100$ to $a_1b_1a_2b_2$; thus the Hardy model is not strongly contextual. However, there is no such path which includes the edge $00$ over $a_1b_1$. This shows the possibilistic contextuality of the model.

Next, we consider the PR box displayed as a bundle on the right of Fig.~\ref{fig:bundle}.
In this case, the model is strongly contextual, and accordingly there is no univocal closed path.
We can see that the PR box is a discrete version of a M\"obius strip.

\section{Contextuality, logic and paradoxes}

We return to our theme that contextuality arguments lie at the borders of paradox, but they do not cross those borders. %
We shall now show that a similar analysis can indeed be applied to some of the fundamental logical paradoxes.

\subsubsection*{Liar cycles} A Liar cycle of length $N$ is a sequence of statements as shown in Fig.~\ref{fig:liar}.
\begin{figure}
\[
\begin{array}{r@{\ \ }l}
S_1 : & S_2 \text{ is true,} \\
S_2 : & S_3 \text{ is true,} \\
\vdots & \\
S_{N-1} : & S_N \text{ is true,} \\
S_N : & S_1 \text{ is false.} \\
\end{array}
\]
\caption{The Liar Cycle}
\label{fig:liar}
\end{figure}

For $N=1$, this is the classic Liar sentence
\[
\begin{array}{r@{\ \ }l}
S : & S \text{ is false.}
\end{array}
\]
These sentences contain two features which go beyond standard logic: references to other sentences, and a truth predicate.
While it would be possible to make a more refined analysis directly modelling these features, we will not pursue this here, noting that it has been argued extensively and rather compellingly in much of the recent literature on the paradoxes that the essential content is preserved by replacing statements with these features by \emph{boolean equations} \citep{levchenkov:00,wen:01,cook:04,walicki:09}. For the Liar cycles, we introduce boolean variables $x_1, \ldots , x_n$, and consider the following equations:
\[ x_1 = x_2, \;\; \ldots , \;\; x_{n-1} = x_n, \;\; x_n = \neg x_1 \Mdot \]
The ``paradoxical'' nature of the original statements is now captured by the inconsistency of these equations.

Note that we can regard each of these equations as fibered over the set of variables which occur in it:
\[
\begin{array}{r@{\ \ }c@{}c@{}c}
\{ x_1, x_2 \} : & x_1 & {} = {} & x_2 \\
\{ x_2, x_3 \} : & x_2 & {} = {} & x_3 \\
\vdots & & & \\
\{ x_{n-1}, x_n \} : & x_{n-1} & {} = {} & x_n \\
\{ x_n , x_1 \} : & x_n & {} = {} & \neg x_1 \\
\end{array}
\]
Any subset of  up to $n-1$ of these equations is consistent; while the whole set is inconsistent.

Up to rearrangement, the Liar cycle of length 4 corresponds exactly to the PR box. The usual reasoning to derive a contradiction from the Liar cycle corresponds precisely to the attempt to find a univocal path in the bundle diagram on the right of Fig.~\ref{fig:bundle}.
To relate the notations, we make the following correspondences between the variables of Fig.~\ref{fig:bundle} and those of the boolean equations:
\[ x_1 \sim a_2, \;\; x_2 \sim b_1, \;\; x_3 \sim a_1, \;\; x_4 \sim b_2  \Mdot \]
Thus we can read the equation $x_1 = x_2$ as ``$a_2$ is correlated with $b_1$'', and $x_4 = \neg x_1$ as ``$a_2$ is anti-correlated with $b_2$''.

Now suppose that we try to set $a_2$ to $1$. Following the path in Fig.~\ref{fig:bundle} on the right leads to the following local propagation of values:
\begin{gather*}
a_2 = 1 \rsqa b_1 = 1 \rsqa a_1 = 1 \rsqa b_2 = 1 \rsqa a_2 = 0 \\
a_2 = 0 \rsqa b_1 = 0 \rsqa a_1 = 0 \rsqa b_2 = 0 \rsqa a_2 = 1
\end{gather*}
The first half of the path corresponds to the usual derivation of a contradiction from the assumption that $S_1$ is true, and the second half to deriving a contradiction from the assumption that $S_1$ is false.

We have discussed a specific case here, but the analysis can be generalised to a large class of examples along the lines of \citep{cook:04,walicki:09}. 

\subsubsection*{The Robinson Consistency Theorem}

As a final remark on the connections between contextuality and logic, we consider a classic result, the Robinson Joint Consistency Theorem \citep{robinson:56}.
It is usually formulated in a first-order setting. The version we will use has the following form: 
\begin{theorem}[Robinson Joint Consistency Theorem]
Let $T_i$ be a theory over the language $L_i$, $i=1,2$. If there is no sentence $\phi$ in $L_1 \cap L_2$ with $T_1 \vdash \phi$ and $T_2 \vdash \neg \phi$, then
$T_1 \cup T_2$ is consistent. 
\end{theorem}
Thus this theorem says that two compatible theories can be glued together. In this binary case, local consistency implies global consistency.
Note, however, that an extension of the theorem beyond the binary case fails. That is, if we have three theories which are pairwise compatible, it need not be the case that they can be glued together consistently. A minimal counter-example is provided at the propositional level by the following ``triangle'':
\[ T_1 = \{ x_1 \lrarr \neg x_2 \}, \; T_2 = \{ x_2 \lrarr \neg x_3 \}, \; T_3 = \{ x_3 \lrarr \neg x_1 \} . \]
This example is well-known in the quantum contextuality literature as the \emph{Specker triangle} \citep{liang:11}. Although not quantum realizable, it serves as a basic template for more complex examples which are.
See \citep{kochen:67,cabello:96,abramsky:11} for further details.

\section{Perspectives and Further Directions}

The aim of this article has been to give an exposition of the basic elements of the sheaf-theoretic approach to contextuality introduced by the present author and Adam Brandenburger in \citep{abramsky:11}, and subsequently developed extensively with a number of collaborators, including Rui Soares Barbosa, Shane Mansfield,  Kohei Kishida, Ray Lal, Carmen Constantin, Nadish de Silva, Giovanni Caru, Linde Wester, Lucien Hardy, Georg Gottlob, Phokion Kolaitis and Mehrnoosh Sadrzadeh.

\subsection{Some further developments in quantum information and foundations}

\begin{itemize}
\item The sheaf-theoretic language allows a unified treatment of non-locality and contextuality, in which results such as Bell's theorem \citep{bell:64} and the Kochen-Specker theorem \citep{kochen:67} fit as instances of more general results concerning obstructions to global sections.
\item A hierarchy of degrees of non-locality or contextuality is identified \citep{abramsky:11}. This explains and generalises the notion of ``inequality-free'' or ``probability-free'' non-locality proofs, and makes a strong connection to logic \citep{abramsky2013rhv}.
This hierarchy is lifted to a novel classification of multipartite entangled states, leading to some striking new results concerning multipartite entanglement, which is currently poorly understood. In joint work with Carmen Constantin and Shenggang Ying, it is shown that, with certain bipartite exceptions, all entangled $n$-qubit states are ``logically contextual'', admitting a Hardy-style contextuality proof. Moreover, the local observables witnessing logical contextuality can be computed from the state \citep{abramsky2016hardy}.

\item The obstructions to global sections witnessing contextuality are characterised in terms of sheaf cohomology  in joint work with Shane Mansfield and Rui Barbosa \citep{abramsky2011cohomology}, and a range of examples are treated in this fashion.
In later work with Shane Mansfield, Rui Barbosa, Kohei Kishida and Ray Lal \citep{AbramskyBKLM15}, the cohomological approach is carried further, and shown to apply to a very general class of ``All-versus-Nothing'' arguments for contextuality, including a large class of quantum examples arising from the stabiliser formalism. 

\item A striking connection between no-signalling models and global sections with signed measures (``negative probabilities'') is established  in joint work with Adam Brandenburger \citep{abramsky:11}. An operational interpretation of such negative probabilities, involving a signed version of the strong law of large numbers, has also been developed  \citep{abramsky2014operational}.
\end{itemize}

\subsection{Logical Bell inequalities}
Bell inequalities are a central technique in quantum information.
The discussion in Section~3 is based on  joint work  with Lucien Hardy \citep{abramsky:12}, in which a general notion of ``logical Bell inequality'', based on purely logical consistency conditions, is introduced, and it is shown that every Bell inequality (\ie every inequality satisfied by the ``local polytope'') is equivalent to a logical Bell inequality. The notion is developed at the level of generality of the sheaf-theoretic framework \citep{abramsky:11}, and hence applies to arbitrary contextuality scenarios, including multipartite Bell scenarios and Kochen-Specker configurations.

\subsection{Contextual semantics in classical computation}

The generality of the approach, in large part based on the use of category-theoretic tools, has been used to show how contextuality phenomena arise in classical computation.

\begin{itemize}
\item An  isomorphism between the basic concepts of quantum contextuality and those of relational database theory is shown in \citep{abramsky2013relational}.
\item  Connections between non-locality and logic have been developed \citep{abramsky2013rhv}. A number of natural complexity and decidability questions are raised in relation to non-locality.
\item Our discussion of the Hardy paradox in Section~5 showed that the key issue was that a local section (assignment of values) could not be extended to a global one consistently with some constraints (the ``support table''). This directly motivated some joint work with Georg Gottlob and Phokion Kolaitis \citep{abramsky:13b}, in which we studied a refined version of \emph{constraint satisfaction}, dubbed ``robust constraint satisfaction'', in which one asks if a partial assignment of a given length can always be extended to a solution. The tractability boundary for this problem is delineated, and this is used to settle one of the complexity questions previously posed in \citep{abramsky2013rhv}.
\item Application of the contextual semantics framework to natural language semantics was initiated in joint work with Mehrnoosh Sadrzadeh \citep{abramsky2014semantic}. In this paper, a basic part of the Discourse Representation Structure framework \citep{kamp1993discourse} is formulated as a presheaf, and the gluing of local sections into global ones is used to represent the resolution of anaphoric references.
\end{itemize}

Other related work includes \citep{abramsky2016possibilities,barbosa2014monogamy,mansfield2013mathematical,constantin2015sheaf,Barbosa:DPhil-thesis,mansfield2014extendability,hyttinen2015quantum,raussendorf2016cohomological,de2015unifying,kishida2016logic}

\subsection*{Envoi}

It has been suggested that complex systems dynamics can emerge most fruitfully at the ``edge of chaos'' \citep{langton1990computation,waldrop1993complexity}.
The range of contextual behaviours and arguments we have studied and shown to have common structure suggest that a rich field of phenomena in logic and information, closely linked to key issues in the foundations of physics, arise at the borders of paradox.

\bibliographystyle{plainnat}
\bibliography{cwpbib}
%,bdbib}
\end{document}